\documentclass[aps,prb,reprint,twocolumn,groupedaddress]{revtex4-2}

\usepackage{graphicx}
\usepackage{bm}
\usepackage{amsmath}  
\usepackage{amssymb}
\usepackage{hyperref}
\usepackage{xcolor}
\usepackage{array}
\usepackage{multirow}
\usepackage{physics}
\usepackage{lipsum}

\begin{document}
\title{Artificial atoms, Wigner molecules, and emergent Kagome lattice in semiconductor moiré superlattices} 
\author{Aidan P. Reddy}
\author{Trithep Devakul}
\author{Liang Fu}
\affiliation{Department of Physics, Massachusetts Institute of Technology, Cambridge, Massachusetts 02139, USA}
\date{\today}
\begin{abstract}
Semiconductor moir\'e superlattices comprise an array of artificial atoms and provide a highly tunable platform for exploring novel electronic phases. We introduce a theoretical framework for studying moir\'e quantum matter that treats intra-moir\'e-atom interactions exactly and is controlled in the limit of large moir\'e period.
We reveal an abundance of new physics arising from strong electron interactions when there are multiple electrons within a moir\'e unit cell. In particular, at filling factor $n=3$, the Coulomb interaction within each three-electron moir\'e atom leads to a three-lobed ``Wigner molecule''. When their size is comparable to the moir\'e period, the Wigner molecules form an emergent Kagome lattice. Our work identifies two universal length scales characterizing the kinetic and interaction energies in moir\'e materials and demonstrates a rich phase diagram due to their interplay.
\end{abstract}
\maketitle
\emph{Introduction ---}
The field of quantum science and engineering has long been fascinated with the creation of artificial atoms and artificial solids with desired properties. Artifical atoms, such as quantum dots and superconducting qubits, exhibit discrete energy levels and provide a physical carrier of quantum information. An array of coupled artificial atoms defines an artificial solid, which may be used for quantum simulation and quantum computing. Recently, the advent of moir\'e materials has provided a remarkably simple and robust realization of artificial solids, offering unprecedented opportunities to explore quantum phases of matter in two dimensions \cite{cao2018unconventional, andrei2020graphene, mak2022semiconductor}.
In particular, moir\'e superlattices of semiconductor transition metal dichalcogenides (TMDs) host strongly interacting electrons in a periodic potential. When the moir\'e period is large, doped electrons are spatially confined to the potential minima, leading to an array of artificial atoms. Electron tunneling between adjacent moir\'e atoms generates moir\'e bands. The charge density can be easily varied across a range of moir\'e band fillings by electrostatic gating, a means of manipulation 
unprecedented in natural solids whose electron density is determined by the chemistry of their constituent atoms. The \emph{in situ} 
tunable atomic number in semiconductor moir\'e materials is a remarkable property.
\par To date, much theoretical analysis has relied on an effective Hubbard model description of interacting electrons in the lowest few moir\'e bands \cite{wu2018hubbard,zhang2020moire, zang2021hartree, morales2022nonlocal}.  
This approach successfully explains and predicts many observed phenomena such as the emergence of Mott insulators at $n=1$ \cite{regan2020mott, tang2020simulation}, incompressible Wigner crystals at fractional fillings $n<1$ \cite{regan2020mott, xu2020correlated, li2021imaging, huang2021correlated, jin2021stripe, padhi2021generalized, morales2022magnetism, zhou2022quantum, Foutty2022} and charge transfer between distinct species of moir\'e atoms at $n>1$ \cite{zhang2020moire, slagle2020charge, zhao2022gate, xu2022tunable, park2022dipole} ($n$ is the number of doped electrons or holes per moir\'e unit cell). However, it is important to note that the characteristic Coulomb interaction energy within a moir\'e atom is often several times larger than the single-particle superlattice gap. As a consequence, the low-energy Hilbert space can be substantially modified by interactions when multiple occupancy of moir\'e atoms is involved. An accurate many-body theory for semiconductor moir\'e systems therefore requires a proper treatment of the intra-moir\'e-atom interaction.
\par
\begin{figure}[ht]
    \centering
\includegraphics[width=\columnwidth]{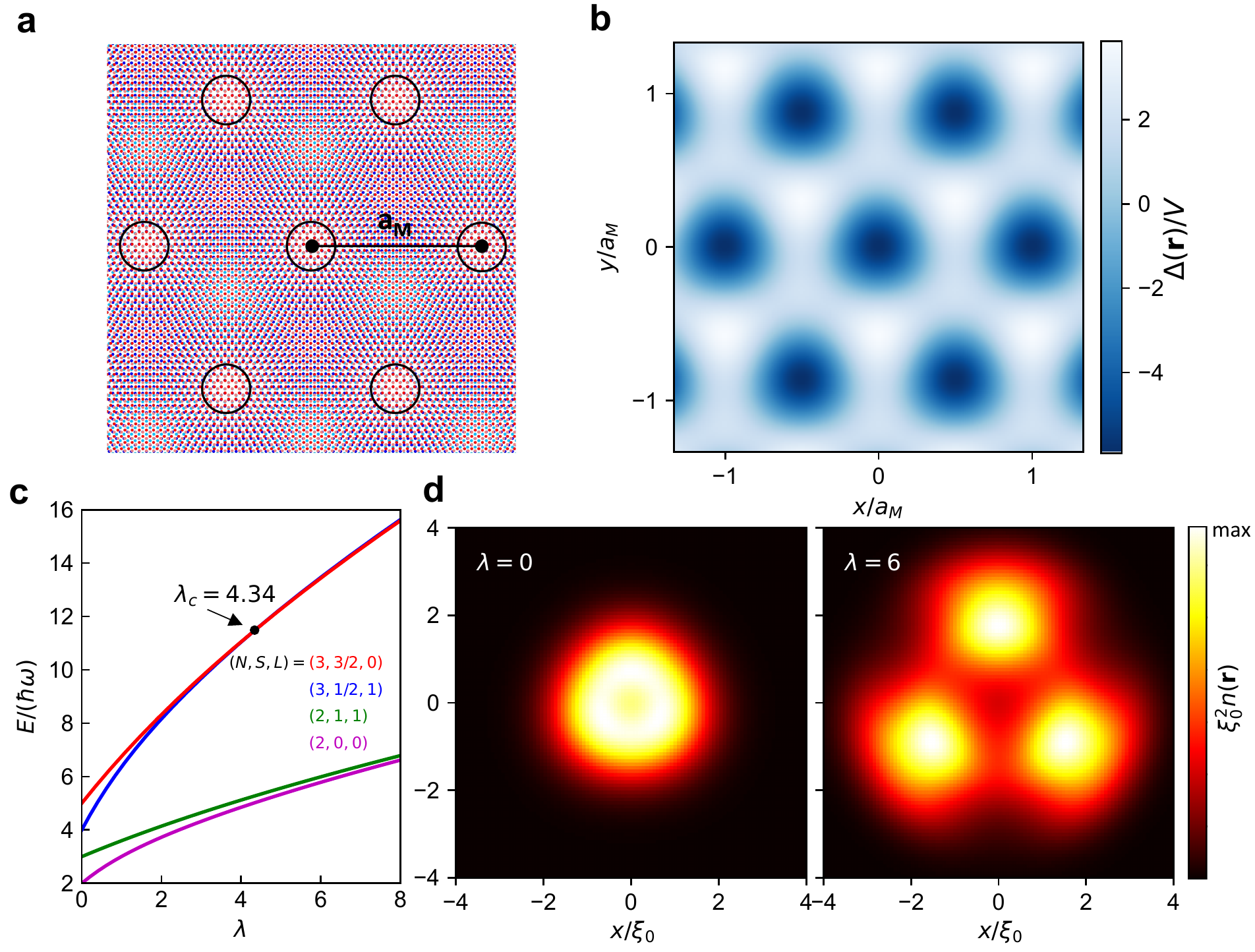}
    \caption{\textbf{Moir\'e atoms and Wigner molecule} (a) Schematic of moir\'e superlattice and (b) corresponding moir\'e potential at $\phi=10^{\circ}$. Its minima, moir\'e atoms, form a triangular lattice. (c) Evolution of each of the high- and low-spin ground states of harmonic helium and lithium (with two and three electrons respectively) with the Coulomb coupling constant $\lambda$. The overall ground state of harmonic lithium transitions from low to high spin at $\lambda_c=4.34$. (d) Charge density distribution of the high spin ground state of moir\'e lithium including a crystal field corresponding to the continuum model parameters ($V=15$meV, $a_M=14$nm, $\phi=10^{\circ}$, $m=0.5m_e$) without (left) and with (right) Coulomb interaction.} 
\label{fig:moireAtomFig}
\end{figure}

 In this work, we predict new physics in semiconductor moir\'e systems arising from interaction effects at higher filling factors.
We first develop an approach to modeling semiconductor moir\'e systems that treats the short-range electronic correlations within a single multi-electron moir\'e atom exactly. 
We model each moir\'e atom as a potential well and solve the interacting few-electron atoms by exact diagonalization. Our approach is controlled in the ``atomic limit'' realized at large moir\'e period where electrons are tightly bound to moir\'e atoms and inter-atomic interactions can be neglected.
For the three-electron atom (moir\'e lithium), we find a distinctive equilateral triangle Wigner molecule charge configuration (see Fig. \ref{fig:moireAtomFig}d), which is stabilized by strong interaction and the threefold anisotropic moir\'e crystal field. 
We further show that when the Wigner molecules' size becomes comparable to the moir\'e period, they collectively form an emergent Kagome lattice of charges at filling factor $n=3$ with electrons localized \emph{between} moir\'e potential minima. This emergent Kagome lattice arises due to the balance between Coulomb interaction and moir\'e potential.
\par The change of charge configuration from triangular or honeycomb to Kagome lattice 
clearly demonstrates that the low-energy Hilbert space of moir\'e systems is strongly filling-dependent due to interaction effects. 
Our work introduces parametrically controlled approximations for treating strong electron-electron interactions in semiconductor moir\'e superlattices and reveals striking consequences of the interplay between quantum kinetic, moir\'e potential, and Coulomb interaction energies.

\par
\emph{Semiconductor moir\'e continuum model---} When two semiconductor TMD monolayers are stacked, a moir\'e pattern appears due to lattice mismatch and/or twist angle. When the moir\'e period $a_M$ is much larger than the monolayer lattice constant, the single-particle moir\'e band structure is accurately described by a continuum model consisting of an effective mass approximation to the semiconductor band edge and a slowly-varying effective periodic potential arising from band edge modulation throughout the moir\'e unit cell. In this work, we always refer to the doped carriers as electrons, regardless of the true sign of their charge. The continuum model Hamiltonian for TMD heterobilayers \cite{wu2018hubbard} (such as WSe$_2$/WS$_2$ and MoSe$_2$/WSe$_2$) and twisted $\Gamma$-valley homobilayers \cite{angeli2021gamma, zhang2021electronic} (such as twisted MoS$_2$) assumes the form
\begin{equation}\label{eq:CMH}
    \mathcal{H}=\frac{\bm{p}^2}{2m} + \Delta(\bm{r})
\end{equation}
where $\Delta(\bm{r})=-2V\sum_{i=1}^{3}\cos(\bm{g}_i\cdot\bm{r}+\phi)$ is an effective moir\'e potential that has the translation symmetry of the superlattice in the first harmonic approximation, $m$ is the effective mass,
and $\bm{g}_i=\frac{4\pi}{\sqrt{3}a_M}(\sin\frac{2\pi i}{3},\cos\frac{2\pi i}{3})$ are the moir\'e reciprocal lattice vectors (Fig. \ref{fig:moireAtomFig}(a)). The minima of $\Delta(\bm{r})$ define a periodic array of moir\'e atoms to which doped charge is tightly bound in the atomic limit which we now examine.

\par 
\emph{Moir\'e atoms ---}
We define an effective Hamiltonian for an electron confined to a moir\'e atom by Taylor expanding $\Delta(\bm{r})$ about the origin:
\begin{align}\label{eq:HMA}
    \begin{split}
        \Delta(\bm{r})\approx \mathrm{const.}+\frac{1}{2}k r^2+c_3\sin{(3\theta})r^3+\ldots
    \end{split}
\end{align}
where $k=16\pi^2 V\cos(\phi)/a_M^2$ and $c_3 = 16\pi^3 V\sin(\phi)/(3^{3/2}a_M^3)$. The result is a circular oscillator with frequency $\omega=\sqrt{k/m}$ along with higher-order, rotation-symmetry-breaking corrections, which we call the moir\'e crystal field. The effective Hamiltonian of an $N$-electron moir\'e atom includes a Coulomb interaction $e^2/(\epsilon |{\bm r}_i -{\bm r}_j|)$ between all of its electron pairs.
\par Both kinetic energy and Coulomb energy favor charge delocalization, whereas the confinement potential favors localization. The characteristic length 
$\xi_{0} \equiv \left( \hbar^2/(mk) \right)^{1/4}$ at which the potential and kinetic energies of a harmonically-confined electron are equal
defines the size of a single-electron moir\'e atom (that is, the extent of its ground state wavefunction). We further introduce the length scale at which the Coulomb and confinement energies of two classical point charges arranged symmetrically about the origin of a harmonic potential are equal, $\xi_{c} = \left(\frac{e^2}{2\epsilon k}\right)^{1/3}$. The ratio of these two length scales is directly related to the dimensionless coupling constant that is the ratio of the intra-atomic Coulomb energy to the atomic level spacing: $\lambda \equiv \frac{e^2/{\epsilon \xi_0}}{\hbar\omega} = 2 (\xi_c/ \xi_0)^3$.

Importantly, the size of the ground state of a few-electron moir\'e atom, which we denote as $\xi$, is on the order of $\xi_0$ for $\lambda \ll 1$ (weak interaction) and $\xi_c$ for $\lambda \gg 1$ (strong interaction), respectively. Therefore for general $\lambda$,  we have $\xi \sim \max\{\xi_0,\xi_c \}$.

\par
\begin{figure}
     \centering
 \includegraphics[width=\columnwidth]{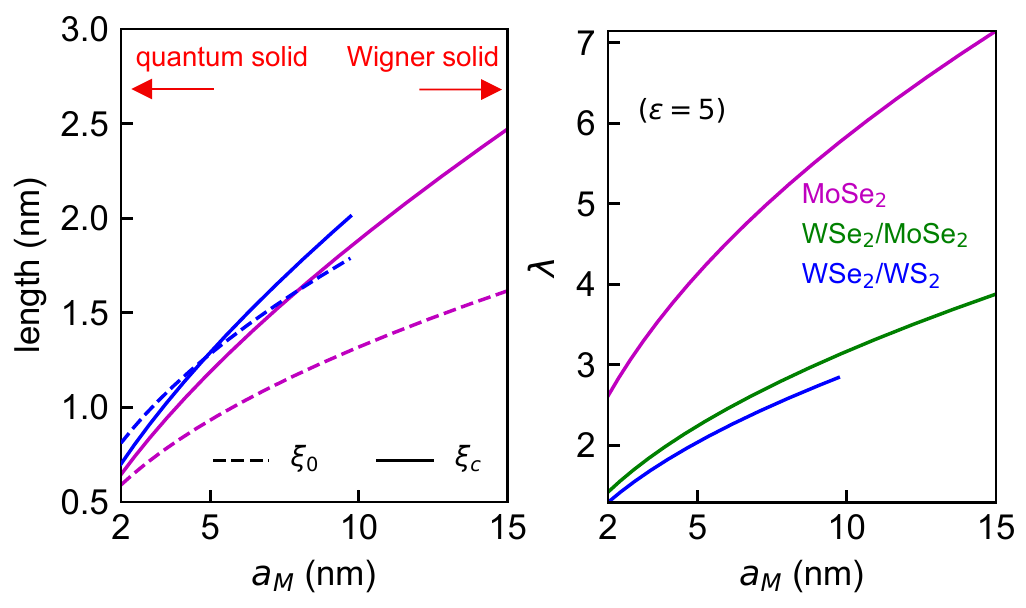}
     \caption{\textbf{Key parameters for TMD moir\'e heterostructures.} Length scales $\xi_0$, $\xi_c$ (left) and coupling constant $\lambda$ (right) in the valence bands of several semiconductor moir\'e systems calculated according to continuum model parameters extracted from density functional theory \cite{zhang2020moire, angeli2021gamma, kometter2022hofstadter}.
 } 
 \label{fig:lengthScales}
 \end{figure}

 Importantly, we observe that the effective spring constant weakens with increasing moir\'e period: $k\propto a_M^{-2}$. It thus follows that
\begin{eqnarray}
\xi_0 \equiv \left(\frac{\hbar^2}{mk}\right)^{1/4}\propto a_M^{1/2}\;;\;\;
\xi_c\equiv\left(\frac{e^2}{2\epsilon k}\right)^{1/3}\propto a_M^{2/3}.
\end{eqnarray}
Consequently, at sufficiently large $a_M$, the hierarchy of length scales $a_M>\xi_c>\xi_0$ is necessarily realized. Then, the size of the few-electron moir\'e atom $\xi$ is parametrically smaller than the distance between adjacent atoms $a_M$ so that intra-atomic Coulomb interaction $\sim e^2/\xi$ dominates over inter-atomic interaction $\sim e^2/a_M$. This self-consistently justifies our treatment of isolated moir\'e atoms as the first step to understanding moir\'e solids. To ground our analysis, we plot the length scales $\xi_c$, $\xi_0$ and the coupling constant $\lambda$ 
as a function of the moir\'e period $a_M$ for three representative TMD heterostructures in Fig. \ref{fig:lengthScales}. 
\par We begin by modeling each moir\'e atom as purely harmonic, neglecting the influence of the crystal field. The single-particle eigenstates of the circular oscillator are labeled by radial and angular momentum quantum numbers $n$ and $l$ with energy $E=(2n+\abs{l}+1)\hbar\omega$. We identify $N=2n+|l|+1$ as the principal quantum number and refer to the circular oscillator eigenstates using electron configuration notation accordingly (i.e. $s$ for $l=0$ and $p$ for $|l|=1$ states).
\par It is known rigorously that the ground state of two electrons with a time-reversal-symmetric Hamiltonian including a symmetric two-body interaction in arbitrary spatial dimensions is a spin singlet \cite{lieb2002theory}. For all interaction strengths, the harmonic helium singlet ground state remains adiabatically connected to the 1s$^2$, single-Slater-determinant ground state at $\lambda=0$, and the triplet first-excited state to the 1s$^1$2p$^1$ state (see Fig. \ref{fig:moireAtomFig}). Although the triplet-singlet energy gap remains positive for all $\lambda$, it asymptotically approaches 0 in the classical limit $\lambda\rightarrow \infty$. 

\par The above theorem does not apply to systems of more than two electrons. In the absence of a Coulomb interaction and moir\'e crystal field, the moir\'e lithium ground state configuration is a 1s$^2$2p spin-doublet with total spin and angular momentum quantum numbers $(S,L)=(1/2,1)$. 
At a critical coupling constant $\lambda_c\approx 4.34$, we find that a ground state level crossing occurs between this low-spin doublet and a high-spin quartet state with $(S,L)=(3/2,0)$ originating from a 1s$^1$2p$^2$ configuration (see Fig. \ref{fig:moireAtomFig})\cite{egger1999crossover, mikhailov2002quantum}. After the level crossing, the energy difference between the low- and high-spin ground states remains small and asymptotes to 0 in the classical limit $\lambda\rightarrow \infty$. Since the energy splittings between the low- and high-spin ground states of the two- and three-electron moiré atoms are small, it should be possible to induce first-order spin transitions with a modest magnetic field \cite{kometter2022hofstadter,ashoori1993n,wagner1992spin}.

\par 
The origin of $S=3/2$ ground state at large coupling constant can be understood from the semiclassical picture. The classical ground state of three interacting electrons in a harmonic potential is an equilateral triangle of side length $\xi_{T}=(2/\sqrt{3})^{1/3}\xi_c$ centered about the origin that spontaneously breaks the rotation symmetry, a configuration which we refer to as a trimer. At finite $\lambda$, quantum fluctuation restores rotational symmetry while preserving the pair correlations of the electron trimer \cite{mikhailov2002quantum}. On the other hand, the moir\'e crystal field term $\sin(3\theta) r^3$ breaks the rotational symmetry explicitly. Since the threefold crystal anisotropy matches with the symmetry of the classical ground state, it stabilizes the triangular ``Wigner molecule" in the presence of a Coulomb interaction.
 As we show in Fig. \ref{fig:moireAtomFig}(c), the charge density of the $(N,S)=(3,3/2)$ state in the absence of the Coulomb interaction ($\lambda=0$) is distorted only mildly by the crystal field, whereas, at $\lambda=6$, it develops a local minimum at the origin and three distinct lobes at the corners of an equilateral triangle. The low-spin state $S=1/2$ exhibits a similar density profile in the presence of the crystal field at moderate and large $\lambda$ (see supplement), again in agreement with the classical limit where spin plays no role.
 The unique charge distribution of the Wigner molecule is a clear consequence of strong interactions and can be directly observed via a local probe such as scanning tunneling microscopy \cite{wach2013charge}.

A heuristic argument for the high-spin $S=3/2$ ground state is that, in the semiclassical expansion, the three-particle exchange processes dominate over that of two-particle exchange processes since the latter must overcome a large Coulomb energy barrier associated with relative coordinates whereas the former need not. Because three-particle exchange amplitudes are necessarily ferromagnetic~\cite{thouless1965exchange, kim2022interstitial}, the ground state at large coupling constant should be fully spin polarized. 


\par \emph{Moir\'e solids---}
Having established the physics of isolated moir\'e atoms, we now turn to their crystalline ensembles: moir\'e solids. We reiterate the important observation that the hierarchy of length scales $a_M > \xi_c > \xi_0$ is necessarily realized at sufficiently large $a_M$. Equivalently, the energy scale of the moir\'e potential depth $V$ necessarily dominates over inter- and intra-atom Coulomb energies $e^2/a_M$, $e^2/\xi$ as well as the quantum zero-point energy $\hbar \omega \propto a_M^{-1}$ associated with harmonic confinement. As a result, the ground state in this regime at integer filling $n$ is an insulating array of $n$-electron moir\'e atoms located at the moir\'e potential minima, which is adiabatically connected to the decoupled limit $a_M\rightarrow \infty$. In this regard, we note that insulating states at integer filling fillings up to $n=8$ have been observed in the conduction band of an MoSe$_2$/WS$_2$ moir\'e superlattice \cite{li2021charge}. 
 
\par
\begin{figure}[h]
    \centering
\includegraphics[width=\columnwidth]{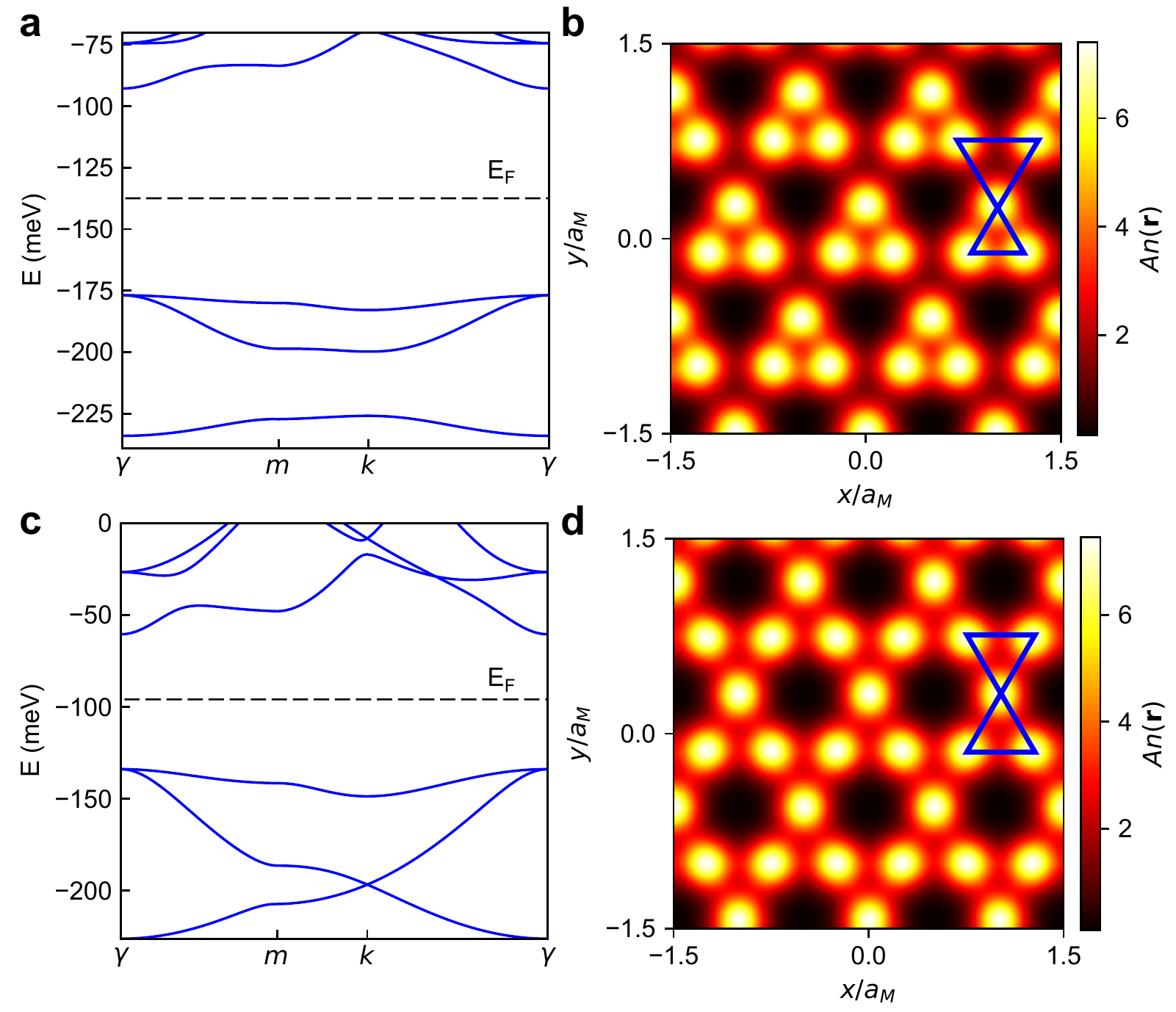}
    \caption{\textbf{Emergent Kagome lattice at $\mathbf{n=3}$} (a) Quasiparticle band structure of self-consistent Hartree-Fock ground state with charge and spin density quantum numbers $(n,s_z)=(3,3/2)$, resembling a Kagome lattice with broken inversion symmetry, i.e. a breathing Kagome lattice. Here we show only the spin-$\uparrow$ bands. (b) Corresponding real space electron density, exhibiting peaks that approximately form a Kagome lattice. $A$ is the moir\'e unit cell area. The parameters used are ($V=15$meV, $\phi=30^{\circ}$, $m=0.5m_e$, $a_M=8$nm, $\epsilon=5$). (c-d) Results for $\phi=60^\circ$ where the continuum model is $D_6$-symmetric and a perfect Kagome lattice emerges at $n=3$.Blue triangles in (b) and (d) indicate triangular Kagome plaquettes.}
\label{fig:emergentKagome}
\end{figure}

 In the following, we investigate the intermediate regime
$a_M\sim \xi_c > \xi_0$ 
and reveal new physics that emerges from inter-atom coupling. 
We apply self-consistent Hartree-Fock theory to the continuum model (Eq. \ref{eq:CMH})\cite{hu2021competing,xie2022valley,pan2022topological}, which treats intra- and inter-atomic interactions on equal footing and accounts for the full periodic moiré potential. Our Hartree-Fock calculation, including the direct inversion of iterative subspace convergence acceleration method, is detailed in the supplement. In particular, at filling factor $n=3$, we find that for realistic model parameters, electrons self-organize into an emergent Kagome lattice (Fig.~\ref{fig:emergentKagome}). This interaction-induced charge order is particularly striking, given that the Kagome lattice sites where electrons localize are saddle points rather than minima of the moir\'e potential.

\par The origin of the emergent Kagome lattice can be understood in the regime $\xi_0\ll \xi_c$ and $a_M$, which is smoothly connected to the classical limit $\xi_0 \rightarrow 0$ or equivalently $m\rightarrow \infty$. 
In this regime, the ground state is determined by the competition between the moir\'e potential and the Coulomb repulsion, which is controlled by the ratio $a_M/\xi_c$. 
As previously discussed, the ground state at large $a_M/\xi_c$ is a lattice of electron trimers of size $\xi_c$ separated by a distance of $a_M$. This charge configuration, which consists of upper and lower triangles of size $\sim \xi_c$ and $a_M$ respectively, can be viewed as a precursor to the Kagome moir\'e solid. As $a_M/\xi_c$ is reduced, the asymmetry in the size of the upper and lower triangles is naturally reduced, so that the charge configuration evolves towards the Kagome lattice. 
 
 \par Our Hartree-Fock calculations fully support the above picture. 
For generic $\phi$,
due to the lack of $D_6$ symmetry, the upper and lower triangles are distinct (Fig.~\ref{fig:emergentKagome}b), thus realizing a breathing Kagome lattice.
Indeed, the corresponding quasiparticle band structure, Fig.~\ref{fig:emergentKagome}a, resembles the Kagome lattice dispersion with broken inversion symmetry. As we show explicitly (see supplement), the three lowest quasiparticle bands are adiabatically connected to the $s$ and $p_x, p_y$ bands on a triangular lattice (as evidenced by the twofold degeneracy at the $\gamma$ point). This fully agrees with our picture of Wigner molecule array evolving into an breathing Kagome lattice. 

\par
\begin{figure}
     \centering
 \includegraphics[width=\columnwidth]{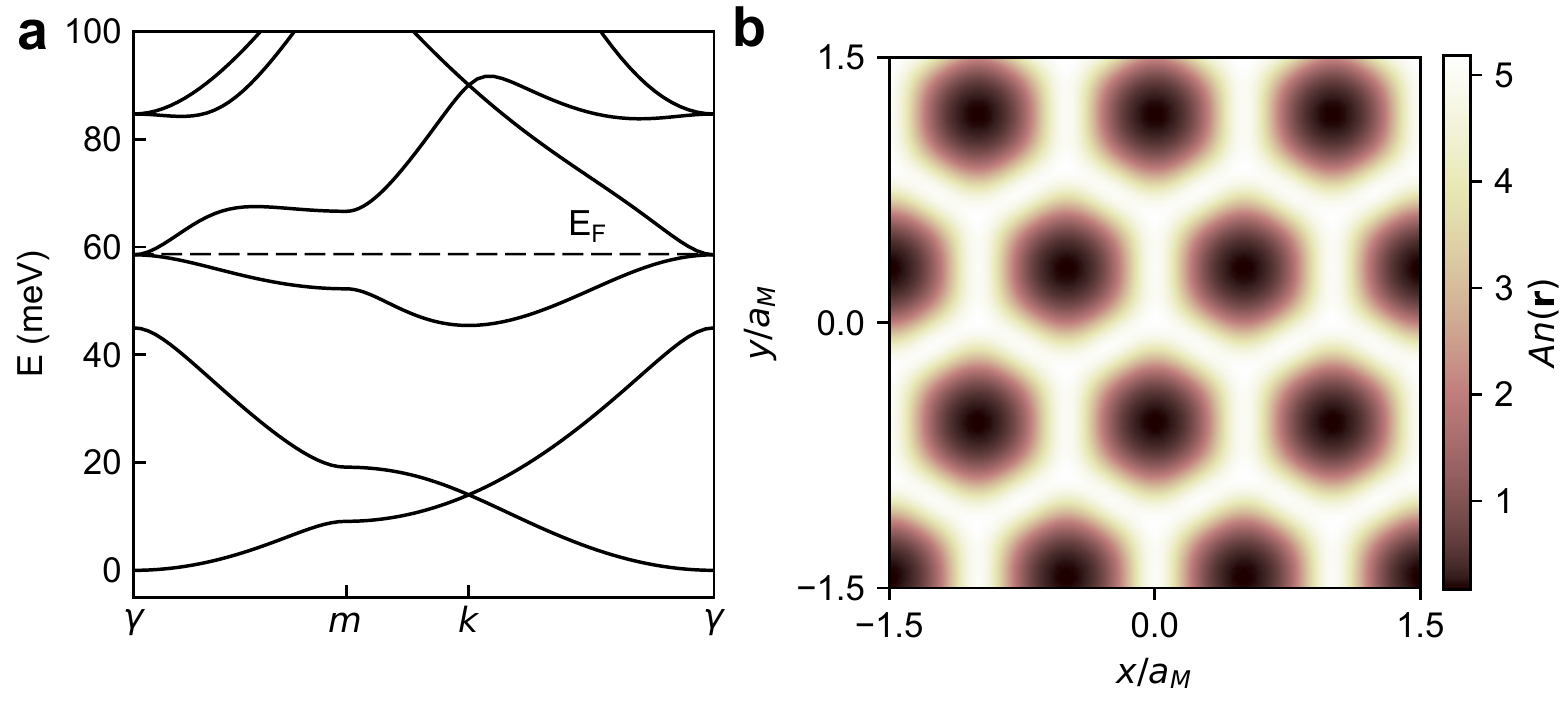}
     \caption{\textbf{Non-interacting bands}. Non-interacting band structure and charge density at $n=3$ for continuum model parameters ($V=15$meV, $\phi=60^{\circ}$, $m=0.5m_e$, $a_M=8$nm, $\epsilon=5$), which contrasts sharply with the interacting case shown in Fig. \ref{fig:emergentKagome}(c,d).
 } 
 \label{fig:honeycombNonInt}
 \end{figure}

 Even more interesting is the case $\phi=60^{\circ}$ as realized in twisted $\Gamma$ valley TMD homobilayers. Here, the underlying moir\'e potential has two degenerate minima per unit cell forming a honeycomb lattice with $D_6$ symmetry. At the filling factor $n=3$, our Hartree-Fock calculation finds that charge assembles into a perfectly symmetric Kagome lattice (Fig.~\ref{fig:emergentKagome}d). 
This is confirmed by the appearance of a Dirac point in the Hartree-Fock band structure (Fig.~\ref{fig:emergentKagome}c). Note that the Kagome band structure found here describes the dispersion of hole quasiparticles in the interaction-induced insulator at $n=3$. 
In contrast, the non-interacting band structure at $\phi=60^{\circ}$, shown in Fig.~\ref{fig:honeycombNonInt}, is gapless at this filling. 
In addition, the Kagome moir\'e solid features a symmetry-protected band degeneracy at $\gamma$ below the Fermi level, which is absent in the noninteracting case.        

\par It is interesting to note that the emergent Kagome lattice of charges minimizes neither the potential energy nor the Coulomb interaction energy. The potential energy favors charges localized at honeycomb lattice sites, while the Coulomb interaction favors a triangular lattice Wigner crystal. Yet, remarkably, our calculations demonstrate that the Kagome lattice emerges as a compromise due to their close competition for realistic material parameters.

\par Although our Hartree-Fock results shown in Fig. \ref{fig:emergentKagome} are for fully spin-polarized electrons, the emergent Kagome insulator at $n=3$ persists regardless of its spin configuration provided that $\xi_c$ is comparable to $a_M$ and sufficiently large compared to $\xi_0$. In this correlated insulating state, the low energy degrees of freedom are localized spins
whose coupling is an interesting problem that we leave to future study.

We now return to the key length scales and coupling constants of real semiconductor moiré systems shown in ~Fig. \ref{fig:lengthScales}. As the coupling constant $\lambda$ increases with $a_M$, the ground state at integer fillings evolves from a quantum solid where quantum effects dominate to a Wigner solid where classical effects dominate. In WSe$_2$/WS$_2$, the change between the two regimes corresponding to $\xi_c=\xi_0$ occurs around $a_M\sim 4.8$nm. In twisted homobilayer MoSe$_2$, we find $\xi_c>\xi_0$ at all values of $a_M$ shown, owing in part to its larger effective mass at the $\Gamma$ valley \cite{angeli2021gamma, zhang2021electronic}. Twisted $\Gamma$ valley homobilayers (WS$_2$, WSe$_2$, MoS$_2$, MoSe$_2$ and MoTe$_2$) also have moiré potentials with an emergent $D_6$ symmetry, 
making them ideal candidates to realize the emergent Kagome solid at small twist angle. 
\par Our demonstration of the emergent Kagome lattice in TMD moir\'e heterostructures paves the way for future investigation of magnetic and topological phases it may host. 
The geometric frustration of antiferromagnetic Heisenberg model on the Kagome lattice makes it a promising candidate for quantum spin liquid \cite{mei2017gapped, mendels2016quantum, broholm2020quantum, motruk2022Kagome, kiese2022tmds}. The prospect that our emergent Kagome lattice may host such a phase deserves further investigation. 


\par \emph{Summary ---}
Our work provides analytically controlled methods to treat strong interaction effect in moir\'e superlattices and r
eveals novel phases of matter at higher filling factors $n>1$. We have identified three key length scales that universally govern the physics of all moir\'e materials: the moir\'e superlattice constant $a_M$, the quantum confinement length $\xi_0$, and crucially, the size of Wigner molecule $\xi_c$. $\xi_0$ characterizes the strength of quantum kinetic energy, and $\xi_c$ the strength of Coulomb interaction.
\par We have established two parameter regions in which theoretical analysis is controlled even for strong interactions. First, when $a_M\gg\xi_0,\xi_c$, moir\'e atoms can be always treated in isolation regardless of the ratio $\xi_c/\xi_0$. By exactly solving the few-electron state of a single moir\'e atom, we have predicted the existence of the Wigner molecule. Second, when $\xi_0\ll\xi_c,a_M$, a self-organized electron lattice is formed to minimize the sum of potential and interaction energy. In particular, we predict an emergent Kagome lattice at the filling $n=3$ for realistic material parameters that correspond to $\xi_c \sim a_M$. More generally, when $\xi_0\ll \xi_c$, the intra-atomic interaction $e^2/(\epsilon \xi_0)$ exceeds the single-particle band gap $\hbar \omega$, hence the effective Hamiltonian cannot be correctly obtained by projecting the interacting continuum model into the lowest moir\'e band. The interplay of the three length scales $a_M$, $\xi_0$ and $\xi_c$, in combination with tunable electron filling $n$, presents a vast phase space and an organizing principle to explore moir\'e quantum matter.
\par
\emph{Acknowledgements---} It is our pleasure to thank Yang Zhang, Faisal Alsallom, and Allan MacDonald for valuable discussions and related collaborations. This work was supported by the Air Force Office of Scientific Research (AFOSR) under award FA9550-22-1-0432.

\nocite{refSMnote}
\nocite{zeng2022strong}
\nocite{laturia2018dielectric}
\nocite{reimann2002electronic}
\nocite{giuliani2005quantum}
\nocite{pulay1980convergence}


%

\end{document}


\title{Supplementary material for \emph{Artificial atoms, Wigner molecules, and emergent Kagome lattice in semiconductor moiré superlattices}}
\author{Aidan P. Reddy}
\author{Trithep Devakul}
\author{Liang Fu}
\affiliation{Department of Physics, Massachusetts Institute of Technology, Cambridge, Massachusetts 02139, USA}
\date{\today}
\maketitle
\tableofcontents
\section{Extended results}
\subsection{Charge density of Wigner molecule}
\begin{figure}[h]
    \centering
\includegraphics[width= \textwidth]{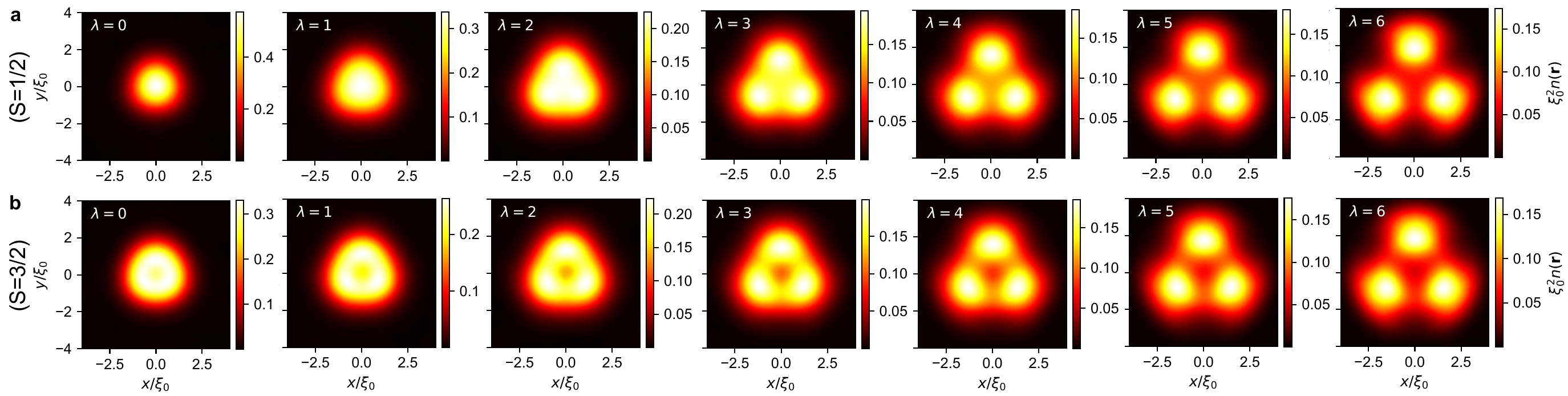}
    \caption{Evolution of the $n=3$, low- (a) and high-spin (b) ground state charge densities with the Coulomb coupling constant $\lambda$. Moir\'e atom parameters are extracted from the continuum model parameters ($a_M=14$nm, $V=15$meV, $m^*=0.5m_e$, $\phi=10^{\circ}$) which are the same as those used in Fig. 1 of the main text.} 
\label{fig:lowHighSpinCharge}
\end{figure}

In Fig. \ref{fig:lowHighSpinCharge}, we plot the charge densities of the low- and high-spin states of moir\'e lithium for various values of the Coulomb coupling constant $\lambda$. At weak interaction, the low-spin state's charge density is significantly more confined than that of the high-spin state. This can be understood most simply in the purely harmonic and isotropic case where the low-spin configuration is $1s^22p^1$ and the high-spin configuration is $1s^12p^2$. Since the $p$ orbitals have a larger radial expectation value, the high-spin state charge density is more diffuse. As $\lambda$ increases, the charge densities of the low- and high-spin states become increasingly similar and both form triangular Wigner molecules. The similarity of the charge density distributions of the two states is expected in classical limit $\lambda\rightarrow \infty$.
\subsection{Wigner molecule solid to emergent Kagome lattice evolution with $\phi$}
\par
\begin{figure}[h]
    \centering
\includegraphics[width= \textwidth]{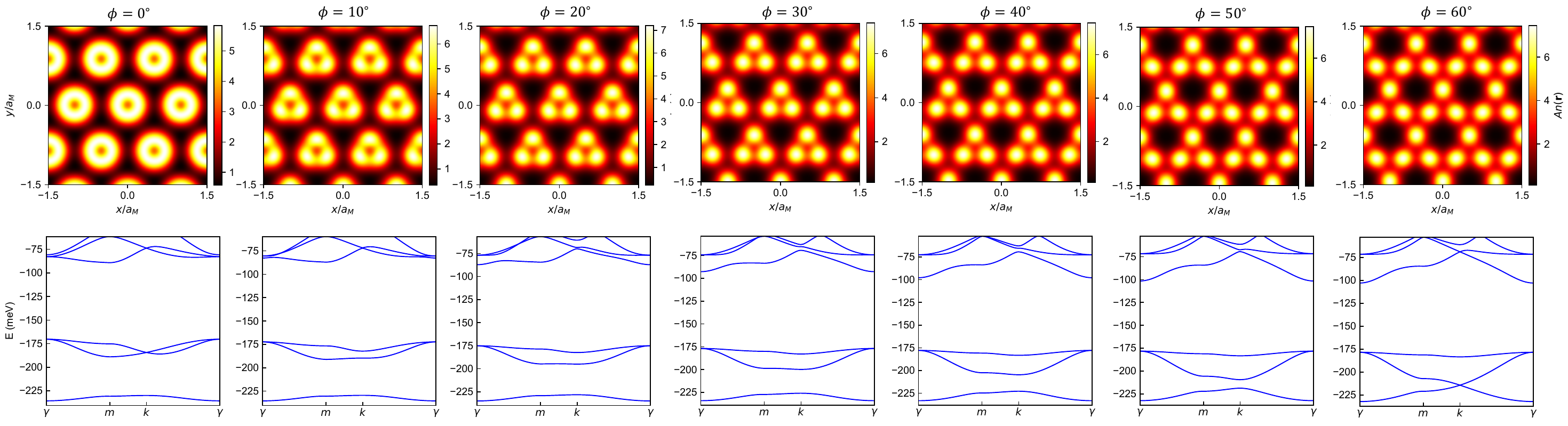}
    \caption{Evolution of the $n=3$, fully-polarized HF ground state bands and charge densities under $\phi: 0^{\circ} \rightarrow 60^{\circ}$. The other continuum model parameters are fixed at ($V=15$meV, $a_M$=8nm, $m^*=0.5m_e$, $\epsilon=5$) (same as in Fig. 3 of main text).} 
\label{fig:WMEKLAdConn}
\end{figure}

 In Fig. \ref{fig:WMEKLAdConn}, we show the charge densities and quasiparticle band structures of fully-spin($=$valley)-polarized electrons at filling factor $n=3$ across a range of values of the continuum model phase parameter $0^{\circ} \leq \phi \leq 60^{\circ}$. When $\phi=0^{\circ}$, the moir\'e potential is $D_6$-symmetric about its minima which comprise a triangular lattice. In this case, the charge density of each moir\'e atom ($\equiv$ potential minimum) is approximately isotropic. For generic $\phi$, the moir\'e potential is only $D_3$-symmetric about its minima. Since in this case the moir\'e atom point group matches that of a classical three-electron equilateral triangle charge configuration, the charge distribution within each moir\'e atom has a substantial triangular anisotropy. When $\phi=60^{\circ}$ the charge triangles at each honeycomb site collectively form a perfect Kagome lattice as discussed in the main text. Our calculations demonstrate that the $\phi=0^{\circ}$ and $\phi=60^{\circ}$ fully-spin-polarized ground states within Hartree-Fock theory are adiabatically connected under the evolution of $\phi$.
\subsection{Kagome to honeycomb transition}
\begin{figure}
    \centering
    \includegraphics[width=0.8\textwidth]{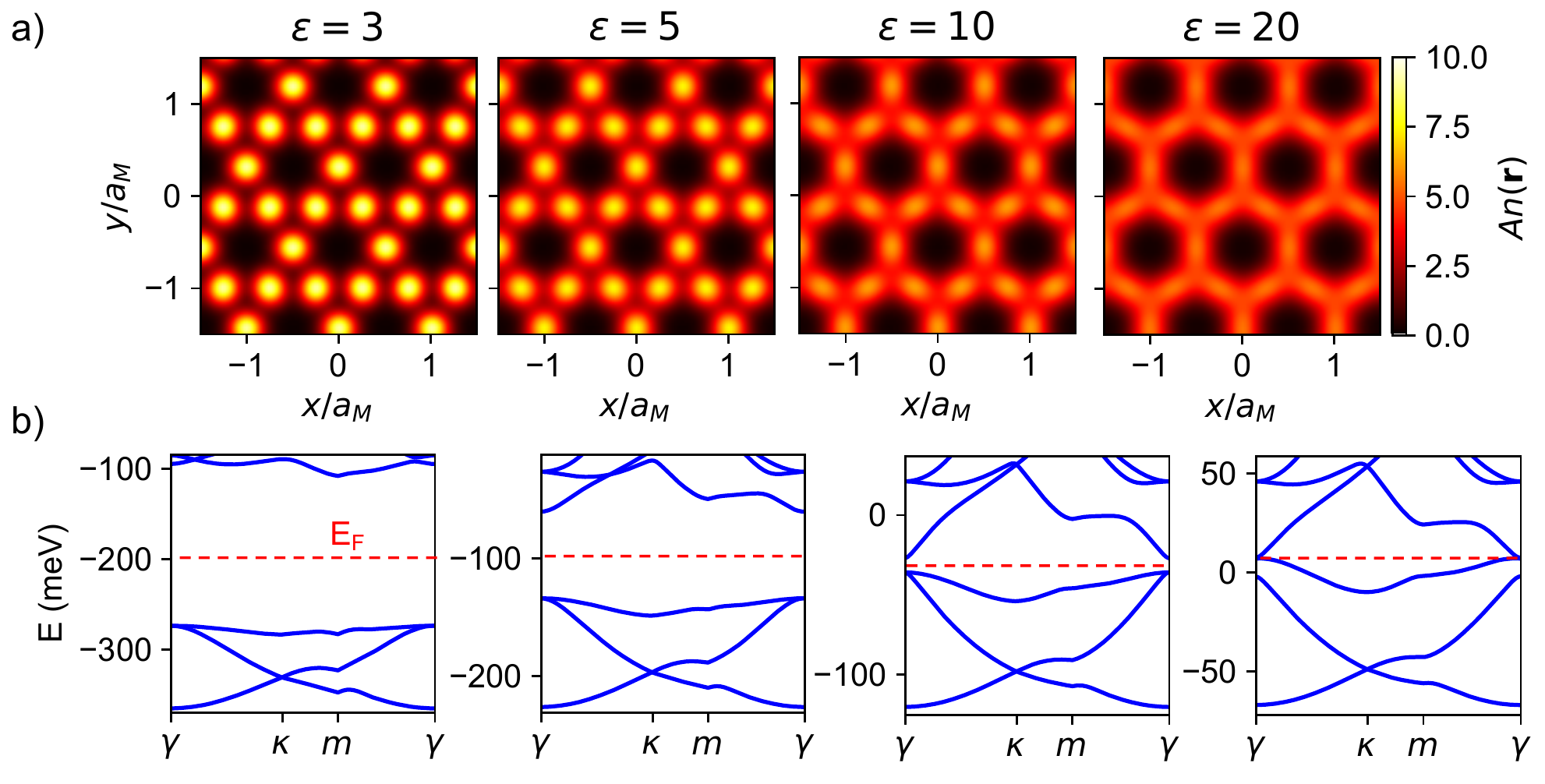}
    \caption{Demonstration of emergent Kagome lattice across broad range of dielectric constants and honeycomb lattice at sufficiently weak interaction. Here we use the same continuum model parameters as in Fig. 3 of the main text: $V=15$meV, $\phi=30^{\circ}$, $m=0.5m_e$, $a_M=8$nm.}
    \label{fig:kagomerobust}
\end{figure}

In Fig.~\ref{fig:kagomerobust}, we study the dependence of the $n=3$ HF ground state on interaction strength, or dielectric constant. The charge densities and quasiparticle band structure show that the emergent Kagome lattice is robust across a broad range of interaction strengths, spanning at least $\epsilon=3-10$. With stronger interactions, the charge density peaks or Kagome lattice sites are more localized and the gap between the highest occupied Kagome band and the next quasiparticle band is large. As interaction strength decreases, the charge peaks become weaker and the gap between the highest kagome band and the next band shrinks. When the interaction strength is sufficiently weak, the quasiparticle band structure resembles the non-interacting band structure and the lowest two occupied quasiparticle bands are of honeycomb lattice character. We remark that the range of interactions strengths over which the emergent Kagome lattice is stable depends on other mode parameters such as the effective mass, moiré period, and moiré potential strength.

We now comment on the effective dielectric constant $\epsilon$. In a medium with in-plane and out-of-plane dielectric constants $\epsilon_{\parallel}$ and $\epsilon_{\perp}$, the Coulomb energy between two point charges in the same plane separated by a distance $r$ is $V(r)=\frac{e^2}{\sqrt{\epsilon_{\parallel}\epsilon_{\perp}} r}$. In bulk hBN, for instance, $\epsilon_{\parallel}=6.93$ and $\epsilon_{\perp}=3.76$, giving $\sqrt{\epsilon_{\parallel}\epsilon_{\perp}}=5.1$ \cite{laturia2018dielectric}. The effective dielectric constant may be treated as a phenomenological parameter and varied from $\sqrt{\epsilon_{\parallel}\epsilon_{\perp}}=5.1$ to account for additional physical effects such as the finite thickness of the TMD layers in an approximate way.

\section{Moir\'e atom calculation}
\subsection{Moir\'e atom Hamiltonian}
The continuum model moir\'e potential in the lowest harmonic approximation is
\begin{equation}
\Delta(\bm{r})=-2V\sum_{i=1}^{3}\cos(\bm{g}_i\cdot\bm{r}+\phi)
\end{equation}
where 
\begin{equation}
    \bm{g}_i=\frac{4\pi}{\sqrt{3}a_M}(\sin\frac{2\pi i}{3},\cos\frac{2\pi i}{3}).
\end{equation}
The first several terms in the Taylor expansion of $\Delta(\bm{r})$ about the origin (a local minimum) in polar coordinates are
\begin{align}
    \begin{split}
        \Delta(r,\theta)/V &= -6\cos(\phi)+8\pi^2\cos(\phi)(r/a_M)^2+\frac{16\pi^3}{3\sqrt{3}}\sin(\phi)\sin(3\theta)(r/a_M)^3-\frac{8\pi^4}{3}\cos(\phi)(r/a_M)^4\\
        &-\frac{16\pi^5}{9\sqrt{3}}\sin(\phi)\sin(3\theta)(r/a_M)^5+\frac{16\pi^6}{405}\cos(\phi)(10-\cos(6\theta))(r/a_M)^6+\ldots
    \end{split}
\end{align}
\par
Let us begin by considering the problem of one electron confined to a minimum of the moir\'e potential. To leading order in $r/a_M$, we can approximate the Hamiltonian as that of a circular harmonic oscillator:
\begin{equation}
    \mathcal{H} = \frac{\bm{p}^2}{2m} + \frac{k}{2}\bm{r}^2 + \ldots
\end{equation}
where $k=16\pi^2 V\cos(\phi)/a_M^2$. 
We now present an algebraic solution for simultaneous energy and angular momentum eigenstates of the circular oscillator. We define the oscillator frequency
\begin{equation}
    \omega = \sqrt{\frac{k}{m}},
\end{equation}
oscillator length
\begin{align}
    \xi_0\equiv \sqrt{\frac{\hbar}{m\omega}},
\end{align}
cartesian ladder operators
\begin{equation}
    a_{x}=\frac{1}{\sqrt{2}}\left(\frac{x}{\xi_0}+\frac{\xi_0p_x}{\hbar}\right),
\end{equation}
and the circular oscillator ladder operators
\begin{align}
    a_{\pm}\equiv \frac{1}{\sqrt{2}}(a_x\mp ia_y).
\end{align}
These circular ladder operators obey the canonical commutation relations
\begin{equation}
    [a_{\pm},a_{\pm}^{\dagger}]=1;\; [a_{\pm},a_{\mp}^{\dag}]=[a_{\pm},a_{\mp}]=[a_{\pm},a_{\pm}]=0
\end{equation}
and the cartesian ladder operators commute analogously. In terms of the circular ladder operators, the Hamiltonian and angular momentum operators are
\begin{align}
    \mathcal{H}=\hbar\omega(a^{\dag}_+a_++a^{\dag}_-a_-+1)
\end{align}
and
\begin{align}
    \begin{split}
    L&\equiv xp_y-yp_x\\
    &=\hbar(a^{\dag}_+a_+-a^{\dag}_-a_-).
    \end{split}
\end{align}
Let us define the ground state $\ket{00}$ according the property
\begin{equation}
    a_{\pm}\ket{00}=0.
\end{equation}
From this definition and the circular ladder operator canonical commutation relations, it follows that
\begin{align}
    \ket{n_+n_-}\equiv \frac{(a_+^{\dag})^{n_+}(a_-^{\dag})^{n_-}}{\sqrt{n_+!n_-!}}\ket{00}
\end{align}
is a simultaneous energy and angular momentum eigenstate:
\begin{align}
    \mathcal{H}\ket{n_+n_-}=(n_++n_-+1)\hbar\omega\ket{n_+n_-} \equiv N\hbar\omega\ket{n_+n_-}
\end{align}
and
\begin{align}
    L\ket{n_+n_-}=(n_+-n_-)\hbar\ket{n_+n_-} \equiv l\hbar\ket{n_+n_-}.
\end{align}
We note that the quantum numbers defined in the main text are $n=\min(n_+,n_-)$, $l=n_+-n_-$, and $N=2n+|l|+1$.
\subsection{Coulomb matrix elements}
Our next task is to calculate the Coulomb matrix elements in this single-particle basis, to which end we follow the approach of Ref. \cite{zeng2022strong}. The Fourier representation of the Coulomb interaction is
\begin{equation}
    U = \int\,\frac{d^2q}{(2\pi)^2}U(\bm{q})e^{i\bm{q}\cdot\bm{r}}
\end{equation}
where
\begin{equation}
    U(\bm{q}) \equiv \frac{2\pi e^2}{\epsilon q}.
\end{equation}
We define the complex variable
\begin{equation}
\tilde{q} \equiv \frac{\xi_0(q_x +iq_y)}{2}
\end{equation}
and the natural energy scale
\begin{equation}
E_c \equiv \frac{e^2}{\epsilon \xi_0} = \lambda\hbar\omega.
\end{equation}
From the definitions of the relative coordinate $\pm$ ladder operators, we have
\begin{equation}
a_{x} = \frac{1}{\sqrt{2}}(a_{+}+ a_{-})
\end{equation}
and
\begin{equation}
a_{y}=\frac{i}{\sqrt{2}}(a_{+}-a_{-}).
\end{equation}
This allows us to express the coordinate operators in terms of the $\pm$ ladder operators as \begin{equation}
x/\xi_0 = \frac{1}{\sqrt{2}}(a_{x}+a^\dagger_{x}) = \frac{1}{2}(a_{+}+a_{-}+a^{\dagger}_{+}+a^{\dagger}_{-})
\end{equation}
and
\begin{equation}
y/\xi_0 = \frac{1}{\sqrt{2}}(a_{y}+a^{\dagger}_{y}) = \frac{i}{2}(a_{+}-a_{-}+a^{\dagger}_{-}-a^{\dagger}_{+}).
\end{equation}
Putting this all together,
\begin{align}
    \begin{split}
        \bm{q}\cdot\bm{r} &= q_xx + q_yy \\
        &=\frac{1}{2}[(q_x+iq_y)(x-iy)+(q_x-iq_y)(x+iy)]\\
        &=\tilde{q}(a_{-}+a_{+}^{\dagger}) + \tilde{q}^*(a_{+} + a_{-}^{\dagger}).
    \end{split}
\end{align}
With this result, we can rewrite the Coulomb matrix element as 
\begin{align}\label{CME1}
    \begin{split}
    U_{ijkl} &= \bra{n_{+i}n_{-i},n_{+j}n_{-j}}U\ket{n_{+k}n_{-k},n_{+l}n_{-l}} \\
    &= \int\frac{d^2q}{(2\pi)^2} U(\bm{q})\bra{n_{+i}n_{-i},n_{+j}n_{-j}}e^{i\bm{q}\cdot({\bm{r_1}-\bm{r}_2})}\ket{n_{+k}n_{-k},n_{+l}n_{-l}}\\
    &= \int\frac{d^2q}{(2\pi)^2} \frac{2\pi e^2}{\epsilon q}\bra{n_{+i}n_{-i}}e^{i[\tilde{q}(a_{-}+a_{+}^{\dagger}) + \tilde{q}^*(a_{+} + a_{-}^{\dagger})]}\ket{n_{+k}n_{-k}}\bra{n_{+j}n_{-j}}e^{-i[\tilde{q}(a_{-}+a_{+}^{\dagger}) + \tilde{q}^*(a_{+} + a_{-}^{\dagger})]}\ket{n_{+l}n_{-l}} \\
    &=2E_c\int\frac{d^2\tilde{q}}{2\pi} \frac{1}{\tilde{q}}\bra{n_{+i}}e^{i(\tilde{q}a_{+}^{\dagger} + \tilde{q}^*a_{+})}\ket{n_{+k}}\bra{n_{-i}}e^{i(\tilde{q}a_{-} + \tilde{q}^*a_{-}^{\dagger})}\ket{n_{-k}}\bra{n_{+j}}e^{-i(\tilde{q}a_{+}^{\dagger} + \tilde{q}^*a_{+})}\ket{n_{+l}}\bra{n_{-j}}e^{-i(\tilde{q}a_{-} + \tilde{q}^*a_{-}^{\dagger})}\ket{n_{-l}}.
    \end{split}
\end{align}
From the Baker-Campbell-Hausdorff formula, we have the identity
\begin{equation}\label{BCH}
    \begin{aligned}
    e^{i(fa^{\dagger}+f^*a)} = e^{-|f|^2/2}e^{ifa^{\dagger}}e^{if^*a},
    \end{aligned}
\end{equation}
which allows us to expand the form factors in Eq. \ref{CME1} as
\begin{align}\label{BCHMatrixElement}
\begin{split}
    \bra{n'}e^{i(fa^{\dagger}+f^*a)}\ket{n} &= e^{-|f|^2/2}\bra{n'}e^{ifa^{\dagger}}e^{if^*a}\ket{n} \\
    &= e^{-|f|^2/2} \sum_{\alpha'=0}^{n'}\sum_{\alpha=0}^{n}\frac{(if)^{\alpha'}(if^*)^{\alpha}}{\alpha'!\alpha!}\bra{n'}(a^{\dagger})^{\alpha'}a^{\alpha}\ket{n} \\
    &= e^{-|f|^2/2} \sum_{\alpha'=0}^{n'}\sum_{\alpha=0}^{n}\frac{(if)^{\alpha'}(if^*)^{\alpha}}{\alpha'!\alpha!}\sqrt{\frac{n'!}{(n'-\alpha')!}}\sqrt{\frac{n!}{(n-\alpha)!}}\delta_{n'-(n-\alpha), \alpha'} \\ 
    &=e^{-|f|^2/2}\sum_{\alpha=0}^{\min(n',n)}\frac{(if)^{n'-(n-\alpha)}(if^*)^{\alpha}}{(n'-(n-\alpha))!\alpha!}\frac{\sqrt{n'!n!}}{(n-\alpha)!} \\
    &= e^{-|f|^2/2} \sum_{\alpha=0}^{\min(n',n)}\frac{i^{n'-n+2\alpha}|f|^{n'-n+2\alpha}(e^{i\phi})^{n'-n}}{(n'-(n-\alpha))!\alpha!}\frac{\sqrt{n'!n!}}{(n-\alpha)!} \\
    &= e^{-|f|^2/2}(ie^{i\phi})^{n'-n}\sum_{\alpha=0}^{\min(n',n)}\frac{(-1)^{\alpha}|f|^{n'-n+2\alpha}}{(n'-(n-\alpha))!\alpha!}\frac{\sqrt{n'!n!}}{(n-\alpha)!} \\
    &= e^{-|f|^2/2}(ie^{i\phi})^{n'-n}(n'!n!)^{1/2}(-1)^{n}\sum_{\alpha=0}^{\text{min}(n',n)}\frac{(-1)^{\alpha}|f|^{n'+n-2\alpha}}{\alpha!(n'-\alpha)!(n-\alpha)!}.
\end{split}
\end{align}
where $f \equiv |f|e^{i\phi}$ and in the final equality we make the substitution $\alpha \rightarrow n-\alpha$. Inserting this into \ref{CME1} yields
\begin{align}
    \begin{split}
        U_{ijkl} &= 2E_c(-1)^{N_j-N_l}(-1)^{N_k+N_l}i^{\Delta N}(n_{i+}n_{j+}n_{k+}n_{l+}n_{i-}n_{j-}n_{k-}n_{l-})^{1/2}\int\,\frac{d^2\tilde{q}}{2\pi}\frac{1}{|\tilde{q}|}e^{-2\tilde{q}^2}e^{i\Delta l \phi} \\
        &\times \left(\sum_{\alpha=0}^{\text{min}(n_{i+},n_{k+})}\frac{(-1)^{\alpha}|\tilde{q}|^{n_{i+}+n_{k+}-2\alpha}}{\alpha!(n_{i+}-\alpha)!(n_{k+}-\alpha)!}\right)\left(\sum_{\beta=0}^{\text{min}(n_{i-},n_{k-})}\frac{(-1)^{\beta}|\tilde{q}|^{n_{i-}+n_{k-}-2\beta}}{\beta!(n_{i-}-\beta)!(n_{k-}-\beta)!}\right) \\
        &\times \left(\sum_{\gamma=0}^{\text{min}(n_{j+},n_{l+})}\frac{(-1)^{\gamma}|\tilde{q}|^{n_{j+}+n_{l+}-2\gamma}}{\gamma!(n_{j+}-\gamma)!(n_{l+}-\gamma)!}\right)\left(\sum_{\eta=0}^{\text{min}(n_{j-},n_{l-})}\frac{(-1)^{\eta}|\tilde{q}|^{n_{j-}+n_{l-}-2\eta}}{\eta!(n_{j-}-\eta)!(n_{l-}-\eta)!}\right) \\
        &= \frac{E_c}{\sqrt{2}}(-1)^{N_j+N_k+\Delta N/2}(n_{i+}n_{j+}n_{k+}n_{l+}n_{i-}n_{j-}n_{k-}n_{l-})^{1/2}\\
        &\times \sum_{\alpha=0}^{\text{min}(n_{i+},n_{k+})}\frac{(-1)^{\alpha}}{\alpha!(n_{i+}-\alpha)!(n_{k+}-\alpha)!} \sum_{\beta=0}^{\text{min}(n_{i-},n_{k-})}\frac{(-1)^{\beta}}{\beta!(n_{i-}-\beta)!(n_{k-}-\beta)!} \\
        &\times \sum_{\gamma=0}^{\text{min}(n_{j+},n_{l+})}\frac{(-1)^{\gamma}}{\gamma!(n_{j+}-\gamma)!(n_{l+}-\gamma)!}\sum_{\eta=0}^{\text{min}(n_{j-},n_{l-})}\frac{(-1)^{\eta}}{\eta!(n_{j-}-\eta)!(n_{l-}-\eta)!}\delta_{\Delta l,0}2^{-p/2}\Gamma\left(\frac{p+1}{2}\right)
    \end{split}
\end{align}
where
\begin{equation}
    \Delta l \equiv (n_{i+} - n_{i-} + n_{j+}-n_{j-})-(n_{k+} - n_{k-} + n_{l+}-n_{l-}),
\end{equation}

\begin{equation}
p \equiv (n_{i+} + n_{k+} - 2\alpha)+(n_{i-} + n_{k-} - 2\beta)+(n_{j+} + n_{l+} - 2\gamma)+(n_{j-} + n_{l-} - 2\eta),
\end{equation}

\begin{equation}
N_i \equiv n_{i+}+n_{i-},
\end{equation}
and

\begin{equation}
    \Delta N = N_i+N_j-N_k-N_l.
\end{equation}
In the last equality, we use the integral identity
\begin{align}
    \begin{split}
    \int\frac{\,d^2\tilde{q}}{2\pi}e^{-2|\tilde{q}|^2}\tilde{q}^{p-1}e^{i\Delta l\phi} &= \int_{0}^{2\pi}\frac{\,d\phi}{2\pi}e^{i\Delta l\phi}\int_0^{\infty}\,d|\tilde{q}|e^{-2|\tilde{q}|^2}\tilde{q}^{p} \\
    &=\delta_{\Delta l,0}2^{-(p+3)/2}\Gamma(\frac{p+1}{2}).
    \end{split}
\end{align}
Additionally, we observe that $\Delta N$ and $\Delta l$ necessarily have the same parity. Since $\Delta l$ must be zero for the matrix element not to vanish by conservation of angular momentum, this implies that $\Delta N$ is always even so $i^{\Delta N} = (-1)^{\Delta N/2}$.
\subsection{Real space charge density and wavefunctions}
\par To compute the real space charge densities of our many-body states, we need to know the wavefunctions of our single-particle states, which are known in the literature as Fock-Darwin states. It is convenient to introduce the complex coordinate and momentum operators
\begin{align}
    z\equiv x+iy
\end{align}
and
\begin{align}
    p_z \equiv \frac{p_x-ip_y}{2}.
\end{align}
In terms of the complex coordinate operators,
\begin{align}
    \begin{split}
        a_-= \frac{z}{2\xi_0}+i\frac{\xi_0p^{\dag}_z}{\hbar}
    \end{split}
\end{align}
and
\begin{align}
    \begin{split}
        a_+= \frac{z^{\dag}}{2\xi_0}+i\frac{\xi_0p_z}{\hbar}.
    \end{split}
\end{align}
Firstly, we verify that
\begin{align}
    \begin{split}
    \psi_{00}&\equiv \bra{z,\bar{z}}\ket{00}\\
    &=\frac{1}{\sqrt{\pi}}e^{-z\bar{z}/(2\xi_0^2)}
    \end{split}
\end{align}
where $z\equiv x+iy$ and $z\equiv x-iy$ is the ground state wavefunction by showing that it is annihilated by $a_+$ and $a_-$:
\begin{align}
    \begin{split}
        \bra{z,\bar{z}}a_+\ket{0,0} = \frac{1}{2\sqrt{\pi}}(\frac{\bar{z}}{\xi_0}-\frac{\bar{z}}{\xi_0})=0.
    \end{split}
\end{align}
\begin{align}
    \begin{split}
        \bra{z,\bar{z}}a_-\ket{0,0} = \frac{1}{2\sqrt{\pi}}(\frac{z}{\xi_0}-\frac{z}{\xi_0})=0.
    \end{split}
\end{align}
The wavefunction of a general state in polar coordinates is 
\begin{equation}
    \psi_{n,l}(r,\theta)=\frac{(-1)^{n}}{\xi_0}\sqrt{\frac{n!}{\pi(n+|l|)!}}e^{il\theta}(r/\xi_0)^{|l|}L_{n}^{|l|}((r/\xi_0)^2)e^{-\frac{(r/\xi_0)^2}{2}}
\end{equation}
where $n\equiv \min(n_+,n_-)$ and $l\equiv n_+-n_-$, as can be verified by acting with the raising operators and converting to polar coordinates.
\subsection{Crystal field}
\par
The high-order terms in the Taylor expansion of the moir\'e potential, which we call the moir\'e crystal field, can be constructed straightforwardly as polynomials in the operator
\begin{equation}
    \frac{re^{i\theta}}{\xi_0}=\frac{x+iy}{\xi_0}=a_{+}+a_{-}^{\dag}
\end{equation}
and its Hermitian conjugate.
\subsection{Exact diagonalization}
\par In all calculations shown, we diagonalize the Hamiltonian in the truncated Hilbert space comprising all Fock states generated from the single-particle states of the circular oscillator in the first 10 shells (i.e. all single-particle states with principal quantum number $N\leq 10$, 55 in total). In purely harmonic calculations without a crystal field, we take advantage of the model's rotation symmetry to block-diagonalize the Hamiltonian in sectors with fixed $L$. In calculations with crystal field, we include crystal field terms through order $r^6$ in order to avoid the divergence of the moir\'e potential to $-\infty$ that would occur upon truncating at lower order and block-diagonalize in sectors with fixed ($L\mod 3$). We always use the model's $SU(2)$ spin rotation symmetry to block-diagonalize in sectors with fixed $S_z$ and infer the total spin quantum number $S$ of a given state by comparing different $S_z$ sectors.
\par Lastly, we note that the electronic structure of multi-electron circular oscillator atoms has been widely investigated  in the context of quantum dots \cite{reimann2002electronic}. What distinguishes our calculation is the inclusion of moiré crystal field, which profoundly affects the three-electron charge distribution, and our elucidation of the relevance of quantum dot physics to strongly interacting electrons in semiconductor moiré superlattices.
\section{Continuum model Hartree-Fock}
\subsection{Theory}
The continuum model Hamiltonian with Coulomb interaction takes the form
\begin{align}
    \begin{split}
        \mathcal{H}&\equiv \sum_{\{\bm{k}\bm{g}s\}}\Big[h_{s'\bm{g'},s\bm{g}}(\bm{k})c^{\dag}_{\bm{k}+\bm{g}'s'}c_{\bm{k}+\bm{g}s}\\
        &+ \frac{1}{2A}\delta_{\bm{k}+\bm{k}'+\bm{g}+\bm{g'},\bm{k}''+\bm{k}'''+\bm{g}''+\bm{g}'''}U(\bm{k}''+\bm{g}''-\bm{k}-\bm{g})c^{\dag}_{\bm{k}''+\bm{g}''s}c^{\dag}_{\bm{k}'''+\bm{g}'''s'}c_{\bm{k}'+\bm{g}'s'}c_{\bm{k}+\bm{g}s}\Big]
    \end{split}
\end{align}
where
\begin{align}
\begin{split}
        h_{s'\bm{g'},s\bm{g}}(\bm{k})&=T+V \\
    &=(\delta_{\bm{g}',\bm{g}}\frac{\hbar^2(\bm{k}+\bm{g})^2}{2m^*}+V_{\bm{g}',\bm{g}})\delta_{s',s}
\end{split}
\end{align}
and
\begin{align}
    U(\bm{q})=\frac{2\pi e^2}{\epsilon q}
\end{align}
where $q\equiv |\bm{q}|$. $\bm{k}$ are moir\'e crystal momenta and $\bm{g}$ are moir\'e reciprocal lattice vectors. We set $U(\bm{0})=0$ to avoid the divergence, thereby assuming a neutralizing uniform charge background. (For this reason, the energies in our HF band plots have a constant offset that depends on $n$ and $\epsilon$.)
\par The first term includes the kinetic energy and moir\'e potential and the second term is the Coulomb interaction. The Hartree-Fock equations can be derived either by enforcing that the linear variation of the energy expectation value of the most general single-Slater-determinant state with respect to the single-particle states it comprises vanishes, or by performing a particle-number-conserving, mean-field decoupling of the Coulomb interaction \cite{giuliani2005quantum}. Here we take the latter approach. The application of Wick's theorem to the Coulomb term yields:
\begin{align}
    \begin{split}
    c^{\dag}_{\bm{k}''+\bm{g}''s}c^{\dag}_{\bm{k}'''+\bm{g}'''s'}c_{\bm{k}'+\bm{g}'s'}c_{\bm{k}+\bm{g}s} &= \langle c^{\dag}_{\bm{k}'''+\bm{g}'''s'}c_{\bm{k}'+\bm{g}'s'} \rangle :c^{\dag}_{\bm{k}''+\bm{g}''s}c_{\bm{k}+\bm{g}s}: + \langle c^{\dag}_{\bm{k}''+\bm{g}''s}c_{\bm{k}+\bm{g}s}\rangle :c^{\dag}_{\bm{k}'''+\bm{g}'''s'}c_{\bm{k}'+\bm{g}'s'}:
    \\ &- \langle c^{\dag}_{\bm{k}''+\bm{g}''s}c_{\bm{k}'+\bm{g}'s'}\rangle :c^{\dag}_{\bm{k}'''+\bm{g}'''s'}c_{\bm{k}+\bm{g}s}: - \langle c^{\dag}_{\bm{k}'''+\bm{g}'''s'}c_{\bm{k}+\bm{g}s}\rangle :c^{\dag}_{\bm{k}''+\bm{g}''s}c_{\bm{k}'+\bm{g}'s'}: + \ldots
    \end{split}
\end{align}
where $\langle \rangle$ denotes an expectation value with respect to a candidate single-Slater-determinant ground state 
\begin{equation}
    \ket{\Psi_{HF}} = \left(\prod_{\bm{k}\alpha \in ({\text{occ}\}}}c^{\dag}_{\bm{k}\alpha}\right)\ket{0},
\end{equation}
``occ" denotes the $N$ eigenstates of the Fock matrix $F$ (which we will define momentarily) with the lowest eigenvalues:
\begin{equation}
    \sum_{s\bm{g}}F_{s'\bm{g'},s\bm{g}}(\bm{k})\psi^{\alpha}_{s\bm{g}}(\bm{k}) = \epsilon_{\bm{k}\alpha}\psi^{\alpha}_{s'\bm{g}'}(\bm{k}),
\end{equation}
and $::$ denotes normal ordering with respect to $\ket{\Psi_{HF}}$. The ellipses represent the fully uncontracted and fully contracted terms, which are ``fluctuations" and constant offsets respectively and which we neglect in HF theory. The HF single-particle eigenvalues $\epsilon_{\bm{k}\alpha}$, which are the quantities that appear in the band structure plots, are equivalent to the change in the energy expectation value upon changing the occupation of the state $\ket{\bm{k}\alpha}$ in $\ket{\Psi_{HF}}$, with the appropriate sign depending on the whether or not the state is occupied in $\ket{\Psi_{HF}}$. This is the statement of Koopmans' theorem. After the summation over spin and momenta, the two terms with plus signs become equivalent as do the two terms with minus signs. The resultant HF potential $U^{HF}$ can be expressed in terms of the single-particle density matrix
\begin{align}
    \rho_{s'\bm{g}',s\bm{g}}(\bm{k})\equiv \langle c^{\dag}_{\bm{k}+\bm{g}s}c_{\bm{k}+\bm{g}'s'}\rangle = \sum_{\alpha \in \{\text{occ}\}}\psi^{\alpha}_{s'\bm{g}'}(\bm{k})\psi^{\alpha*}_{s\bm{g}}(\bm{k})
\end{align}
as
\begin{align}
    \begin{split}
        U^{HF}_{s'\bm{g}',s\bm{g}}(\bm{k}) &= J+ K \\
        &= \frac{1}{A}\sum_{\bm{g}''\bm{k}'}[\delta_{s',s}\sum_{s''}\rho_{s''\bm{g}'+\bm{g}'',s''\bm{g}+\bm{g}''}(\bm{k}')U(\bm{g}'-\bm{g})-\rho_{s'\bm{g}'+\bm{g}'',s\bm{g}+\bm{g}''}(\bm{k}')U(\bm{g}''+\bm{k}'-\bm{k})].
    \end{split}
\end{align}
The HF effective single-particle Hamiltonian is
\begin{equation}
        \mathcal{H}_{HF}\equiv \sum_{\{\bm{k}\bm{g}s\}}F_{s'\bm{g'},s\bm{g}}(\bm{k})c^{\dag}_{\bm{k}+\bm{g}'s'}c_{\bm{k}+\bm{g}s}\\
\end{equation}
where
\begin{align}
    \begin{split}
    F_{s'\bm{g'},s\bm{g}}(\bm{k}) &= T+V+J+K\\
    &= h_{s'\bm{g'},s\bm{g}}(\bm{k}) + U^{HF}_{s'\bm{g}',s\bm{g}}(\bm{k})
    \end{split}
\end{align}
is the Fock matrix. We note that the total energy of the self-consistent HF ground state is
\begin{align}
\begin{split}
        E &\equiv \bra{\Psi_{HF}}\mathcal{H}\ket{\Psi_{HF}}\\
    &= \bra{\Psi_{HF}}T+V+\frac{1}{2}(J+K)\ket{\Psi_{HF}} \\
    & =\Tr\left(\rho\left[T+V+\frac{1}{2}(J+K)\right]\right) \\
    &= \sum_{\{\bm{k}\bm{g}s\}}\rho_{s'\bm{g'},s\bm{g}}(\bm{k})\left[h_{s\bm{g}, s'\bm{g'}}(\bm{k})+\frac{1}{2}U^{HF}_{s\bm{g}, s'\bm{g}'}(\bm{k})\right].
\end{split}
\end{align}
The factor of $1/2$ is necessary to avoid double-counting the pairwise Coulomb interaction energies.
In the calculations presented in this work, we have enforced that the HF solution have the same space group symmetry as the single-particle moir\'e potential. Additionally, we have used ``unrestricted" HF theory, which amounts to enforcing that the single-particle density matrix be diagonal in its $S_z$ indices: $\rho_{s'\bm{g'},s\bm{g}}(\bm{k})\propto \delta_{s',s}$ but not enforcing that $\rho_{\uparrow\bm{g'},\uparrow\bm{g}}(\bm{k})=\rho_{\downarrow\bm{g'},\downarrow\bm{g}}(\bm{k})$ (``restricted" HF). Moreover, since we have only studied fully-spin-polarized states, $\rho_{s'\bm{g'},s\bm{g}}(\bm{k})\propto \delta_{s',s}\delta_{s,\uparrow}$.

\par We find a self-consistent solution iteratively starting from an ansatz single-particle density matrix. For the calculations shown, we use the single-particle density matrix corresponding to the ground state in the absence of Coulomb interaction with some random noise added as our ansatz. Let us define the $\rho_i$ to the be input density matrix of the $i^{th}$ iterative step and $F_i=F(\rho_i)$ to be the output Fock matrix of the same step. In the naive iterative method, the input density matrix of the $(i+1)^{th}$ step is $\rho_{i+1}\equiv \rho(F_i)$. The self-consistency condition is that the single-particle density matrix computed according to the $i^{th}$ Fock matrix be the same as the single-particle density matrix from which the $i^{th}$ Fock matrix was computed:
\begin{equation}
    \rho_{i+1}=\rho(F(\rho_i))=\rho_i.
\end{equation}
We measure the degree to which this condition is satisfied through the error matrix
\begin{equation}
\Delta_i =\rho_{i+1}-\rho_i.
\end{equation}
In particular we check the quantities
\begin{equation}
    e^{(1)}_i \equiv \Tr \left(\Delta_i^{\dag}\Delta_i\right) =\sum_{\{\bm{k}\bm{g}s\}}\Delta^*_{s'\bm{g}',s\bm{g}}(\bm{k})\Delta_{s'\bm{g}',s\bm{g}}(\bm{k})
\end{equation}
and
\begin{equation}
    e^{(2)}_i \equiv \max\{|\Delta_i|\}.
\end{equation}
For the calculations shown, we obtain $e^{(1)}<10^{-13}$ and $e^{(2)}<10^{-7}$. We also check that the relative change in the energy value $(\Delta E)_i = (E_i - E_{i-1})/E_{i}<10^{-7}$.
\subsection{Convergence acceleration via direct inversion of iterative subspace (DIIS)}
To accelerate the convergence of this self-consistency loop, we employ a version of the ``direct inversion of iterative subspace" (DIIS) method, also known as Pulay mixing \cite{pulay1980convergence}. The essence of this method is that instead of using as the input density matrix of the $i^{th}$ iterative step the output density matrix of the previous ($\rho_i=F(\rho_{i-1})$), we use a linear combination of the density matrices of the previous $m$ iterations
\begin{equation}\label{DIIS1}
    \rho_i = \sum_{j=i-m}^{i-1}c_j\rho_j.
\end{equation}
such that the magnitude of the extrapolated error matrix $\Delta$ is minimized. If $\rho_i$ is the density matrix of an $N$-particle state, its trace must be equal to $N$, placing a constrain on the $c_j$:
\begin{equation}\label{constraint}
    \Tr\rho_i=N \iff \sum_{j}c_j=1.
\end{equation}
If we assume that the changes in the input and output density matrices over the previous several iterations are small (i.e. if $\bar{\rho}\equiv (\sum_j \rho_j)/m$, then $\delta\rho_j\equiv \rho_j-\bar{\rho}$ is small) then:
\begin{align}
\begin{split}
    \Delta_i &\equiv \Delta(\rho_i) \\
    &= \rho\left(F\left(\rho_i\right)\right) - \rho_i \\
    &= \rho(F(\sum_{j}c_j\rho_j)) - \sum_{j}c_j\rho_j \\
    &= \rho(F(\sum_{j}c_j\left(\bar{\rho}+\delta\rho_j\right))) - \sum_{j}c_j\rho_j \\
    &\approx  \rho(F(\sum_{j}c_j\bar{\rho}))+\sum_jc_j\rho(F(\delta\rho_j)) - \sum_{j}c_j\rho_j \\
    &\approx \sum_j c_j \left(\rho(F(\rho_j)) -\rho_j\right) \\
    &= \sum_{j=i-m}^{i-1} c_j \Delta(\rho_j)
\end{split}
\end{align}
and we have used Eq. \ref{constraint} in between the third and second last lines. We have approximated $\rho(F(\rho'))$ as linear to leading order in variations of $\rho'$. Therefore, the error matrix of $\rho_i$ is approximately equal to the corresponding linear combination of the error matrices of the previous iterations.
The extrapolated error matrix norm is
\begin{align}
\begin{split}
        e^{(1)}_i &= \Tr\left(\Delta_i^{\dag}\Delta_i\right) \\
&=\sum_{j}\sum_{k}c_{j}^{*}c_k\Tr\left(\Delta^{\dag}_j\Delta_k\right) \\
    &\equiv \sum_{j}\sum_{k}c_{j}^{*}c_kB_{jk}.
\end{split}
\end{align}
This quantity can be minimized with the method of Lagrange multipliers:
\begin{equation}
    \mathcal{L}(\rho_i) =\Tr\left(\Delta_i^{\dag}\Delta_i\right) + \lambda(1-\sum_{j}c^{*}_j)
\end{equation}
\begin{equation}
    \partial_{c_j^{*}}\mathcal{L}(\rho_i) = \sum_kc_kB_{jk} - \lambda =0
\end{equation}
Here we have used the fact that $\sum_{j}c_j=1 \rightarrow \sum_{j}c_j^*=1$. In matrix form, this equation is
\begin{equation}
\begin{pmatrix}
    B & 1 \\
    1 & 0
\end{pmatrix}\begin{pmatrix}
\bm{c} \\
-\lambda
\end{pmatrix}=
\begin{pmatrix}
0 \\
1
\end{pmatrix}.
\end{equation}
$\rho_i$ is determined by the $c_j$ that solve this matrix equation, hence ``direct inversion of iterative subspace".
\par In practice, we keep the previous three iterations in the iterative subspace, $m$=3. We also use
\begin{equation}
    \rho_i = \sum_{j=i-m}^{i-1}c_j(\rho_j + \beta \Delta_j)
\end{equation}
instead of Eq. \ref{DIIS1}. It is necessary to introduce with some weight the error matrix of each iteration because otherwise the iterative subspace does not evolve. Anecdotally, we find that setting $\beta=1$ generally works well, which is equivalent to using Eq. \ref{DIIS1} but with $\rho_j \rightarrow \rho(F(\rho_j))$. We note that a single-particle density matrix $\rho$ corresponds to a single-Slater-determinant state if and only of $\rho^2 = \rho$. From this condition, it is clear that a linear combination of single-particle density matrices of distinct single-Slater-determinant states does not correspond to a single-Slater-determinant state itself because it satisfies the inequality $\Tr\rho^2< \Tr\rho$. For this reason, an arbitrary $\rho_i$ in DIIS does not strictly correspond to a single-Slater-determinant. However, when $\rho_i$ satisfies the self-consistency condition, it necessarily corresponds to a single-Slater-determinant. So, while it searches beyond the space of single-Slater-determinant states during intermediate iterations, DIIS always returns a single-Slater-determinant state upon self-consistency.
\subsection{Momentum space sampling}
To construct our plane wave basis, we must define a finite ``mesh" of crystal momenta $\bm{k}$ that evenly sample a momentum space unit cell and a finite reciprocal lattice of vectors $\bm{g}$. Together, these objects define a set of plane waves $\ket{\bm{k}+\bm{g}}$ that tile momentum space in a way that respects the point group symmetry of the moir\'e potential, at least within a high-energy cutoff. The density of the mesh defines the infrared cutoff and the largest shell included in the reciprocal lattice defines the ultraviolet cutoff of our momentum space sampling.
\par
For a triangular lattice, we can make the choice of real-space lattice basis vectors
\begin{equation}
    \begin{aligned}
    \bm{a}_1=a_M(-\frac{1}{2},\frac{\sqrt{3}}{2}); \\
    \bm{a}_2=a_M(1, 0)
\end{aligned}
\end{equation}
and reciprocal lattice basis vectors
\begin{equation}
    \begin{aligned}
    \bm{b}_1=\frac{4\pi}{\sqrt{3}a_M}(0, 1); \\
    \bm{b}_2=\frac{4\pi}{\sqrt{3}a_M}(\frac{\sqrt{3}}{2},\frac{1}{2}).
\end{aligned}
\end{equation}
The Monkhorst-Pack algorithm is a standard method for generating the mesh. It produces an $N\times N$ rhombohedral mesh centered about the origin. A single mesh point is defined by
\begin{equation}
    \bm{k}_{ij}=\frac{2i-N-1}{2N}\bm{b}_1+\frac{2j-N-1}{2N}\bm{b}_2
\end{equation}
and there is one mesh point for every $(i,j)\in \{1,N+1\}^2$.
\begin{figure}[h]
    \centering
\includegraphics[width=0.5\textwidth]{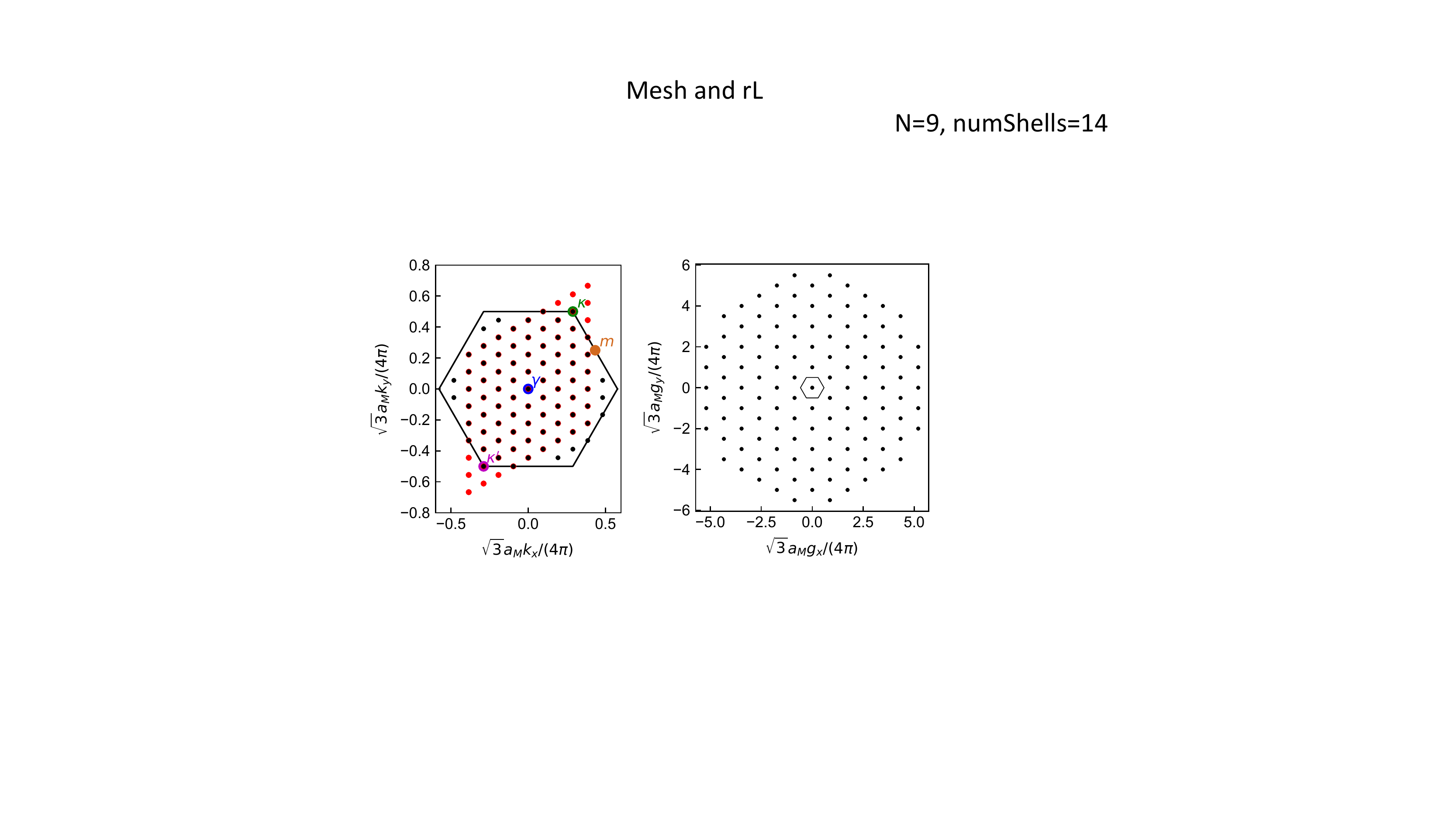}
    \caption{$9\times9$ space $k$-space mesh (left) and 14-shell truncated reciprocal lattice (right) used in continuum model HF calculation. Black hexagon is border of first Brillouin zone. Black points in the left plot are first-Brillouin-zone equivalents of the rhombohedral mesh points in red that were used in the calculation.} 
\label{fig:meshRL}
\end{figure}
\par The reciprocal lattice can be generated first by producing the set of points $\bm{g}_{ij}$ in the same way except without the $2N$ denominator. The resultant reciprocal lattice is a rhombus, and therefore includes some reciprocal lattice vectors of larger magnitude than others than it includes and does not have the rotation symmetry of the full reciprocal lattice. To resolve these shortcomings, we further refine this reciprocal lattice by identifying the $n$ lowest unique $|\bm{g}|$ in the initial mesh and then removing the $\bm{g}$'s that have the largest magnitude. Doing so produces a mesh of the lowest $n$ shells.
\par The dimensionality of the Hilbert space at each crystal momentum is equal to the number of reciprocal lattice vectors in the mesh. Therefore, the number of bands is also equal to the number of reciprocal lattice vectors. We explicitly show our momentum space sampling in Fig. \ref{fig:meshRL}. We emphasize that the emergent Kagome lattice would be impossible to see in a Hartree-Fock calculation in which the single-particle Hilbert space is projected to the lowest several moir\'e bands.
\par Lastly, we note that, to produce the band structure plots shown throughout this work, we interpolate the Hartree-Fock potential matrix $U^{HF}_{s'\bm{g}',s\bm{g}}(\bm{k})$ along a continuous line through momentum space from the mesh points at which we explicitly calculate it.


%